# Federated Data Analytics for Cancer Immunotherapy: A Privacy-Preserving Collaborative Platform for Patient Management

MIRA, RAHEEM

Faculty of Computers & Artificial Intelligence, Cairo University, Cairo, Egypt.

Scientific Academy for Service Technology e.V. (ServTech), Potsdam, Germany

MICHAEL, PAPAZOGLOU

Scientific Academy for Service Technology e.V. (ServTech), Potsdam, Germany

University of New South Wales, Sydney, Australia

University of Macquarie, Sydney, Australia

BERND, KRÄMER

Scientific Academy for Service Technology e.V. (ServTech), Potsdam, Germany

NEAMAT, EL-TAZI

Faculty of Computers & Artificial Intelligence, Cairo University, Cairo, Egypt

AMAL, ELGAMMAL

Egypt University of Informatics, New Administrative Capital, Cairo, Egypt.

Faculty of Computers & Artificial Intelligence, Cairo University, Cairo, Egypt

Scientific Academy for Service Technology e.V. (ServTech), Potsdam, Germany

Connected health is a multidisciplinary approach focused on health management, prioritizing patient needs in the creation of tools, services, and treatments. This paradigm ensures proactive and efficient care by facilitating the timely exchange of accurate patient information among all stakeholders in the care continuum. The rise of digital technologies and process innovations promises to enhance connected health by integrating various healthcare data sources. This integration aims to personalize care, predict health outcomes, and streamline patient management, though challenges remain, particularly in data architecture, application interoperability, and security. Data analytics can provide critical insights for informed decision-making and health co-creation, but solutions must prioritize end-users, including patients and healthcare professionals. This perspective was explored through an agile System Development Lifecycle in an EU-funded project aimed at developing an integrated AI-generated solution for managing cancer patients undergoing immunotherapy. This paper contributes with a collaborative digital framework integrating stakeholders across the care continuum, leveraging federated big data analytics and artificial intelligence for improved decision-making while ensuring privacy. Analytical capabilities, such as treatment recommendations and adverse event predictions, were validated using real-life data, achieving 70%-90% accuracy in a pilot study with the medical partners, demonstrating the framework's effectiveness.

## 1 INTRODUCTION

A new wave of digital technologies — including wearables, big data, cloud computing, simulations, Artificial Intelligence (AI), and Machine Learning (ML)—is evolving and transforming traditional healthcare. Connected health refers to a conceptual model for health management in which digital technologies, including tools and services, are created with the requirements of the patient in mind, ensuring that the patient receives care in the most proactive and effective manner [1], [2]. Connected health enables all stakeholders in the care continuum to be connected through timely exchange and presentation of accurate and precise patient information. Connected health offers boundless potential to address many challenges inherent in traditional medical systems. Key benefits include: (i) fostering patient-centered collaboration among stakeholders across the care continuum—such as doctors, patients, and caregivers—enabling seamless integration and driving digital transformation; (ii) enhancing the monitoring of patient behaviors beyond conventional care settings, encompassing areas like medication adherence, and tailored interventions; and (iii) reducing costs through improved care coordination and operational efficiencies, achieved by gaining deeper insights into patient activities. This minimizes unnecessary service utilization and optimizes resource allocation for better outcomes.

Connected health holds immense promise, but numerous technical and scientific challenges must be addressed before it can achieve its full potential. Although much has been published recently in this field, R&D must advance in the most meaningful manner leading to value-added connected health offerings [3]. In particular:

1. Architectural design, applications and interoperability pose a real challenge. New architectural designs for connected health applications need to exploit the rapid advancement in ICT, including cloud computing and edge computing. Furthermore, Big data challenges in healthcare are further complicated by the fragmentation and dispersion of heterogeneous data across stakeholders. Clinical data often remains siloed, creating a disjointed view of patient records and reducing the efficiency and consistency of treatment delivery. Addressing these issues requires robust solutions for data integration, enabling the linkage of diverse datasets.

2. Security and privacy in connected health is a major issue, as systems rely on internet connectivity. To ensure compliance with regulatory bodies, such as GDPR[1] and HIPAA[2], healthcare systems should impose guaranteed security and privacy methods.
3. Data analytics has immerse potential by applying machine learning, deep learning and AI techniques to draw personalized insights and recommendations. For example, predicting/tracking of adverse events and drug efficacy. However, solutions must be developed by taking a patient-centric approach. Ensuring that solutions address well-defined healthcare challenges is critical to their success.

As part of the H2020 QUALITOP[3] project, we have conducted a comprehensive in-depth analysis of requirements of managing cancer patients receiving immunotherapy by focusing on the three aforementioned challenges. These requirements are conceptualized as a smart digital framework and iteratively validated by medical partners. This partnership ensured that the developed solutions are not only theoretically sound but also practically applicable, addressing the real needs and challenges faced by healthcare professionals. We concluded that a successful implementation and deployment of a connected smart healthcare system requires an agile, robust, reliable, secure, and scalable platform designed to adhere to healthcare data standards and facilitate seamless information exchange. Built on this foundation, we have developed the Smart Digital Health Platform (SDHP) that offers collaborative, customizable querying, and data analytics tools to address the diverse needs of stakeholders. The SDHP prioritizes data protection, security, and confidentiality by incorporating guaranteed security and privacy methods and procedures. This article contributes by a consolidated connected healthcare system that addresses the three challenges highlighted above. In particular:

1. The design and implementation of the SDHP that provides the required collaborative and analytical tools for patient-centered self-management. The SDHP platform realizes the reference architecture for a smart digital platform as proposed in [4].
2. The SDHP leverages an ontology-driven approach to tackle the challenges of syntactic, schematic and semantic heterogeneity in big data. This ontology-based semantic layer serves as a unifying foundation for harmonizing disparate medical data sources, enabling seamless integration and allowing for more comprehensive and accurate analysis.
3. Analytical query patterns, consisting of recurring patterns of retrieval and analytical queries, are iteratively gathered in collaboration with medical partners. These analytical capabilities are implemented on patients receiving CAR T-cell or immune checkpoint inhibitors in real-life in four EU countries: France, The Netherlands, Portugal and Spain, which surmount the known limitations of Randomized Clinical Trails (RCTs) [5]). A federated data analytics approach is exploited, ensuring patient privacy while facilitating seamless data analysis.

---

1 General Data Protection Regulation: https://gdpr.eu
2 Health Insurance Portability and Accountability Act: https://www.ncbi.nlm.nih.gov/books/NBK500019/
3 Monitoring multidimensional aspects of QUAlity of Life after cancer ImmunoTherapy - an Open smart digital Platform for personalized prevention and patient management: https://cordis.europa.eu/project/id/875171/reporting

The validity, efficacy and applicability of the SDHP are ascertained by its full implementation and its iterative validation and evaluation by the medical partners and patient representatives. Analytical query patterns implementation showed an approximate accuracy of 70-90%.

The remainder of this paper is structured as follows: related work is discussed in Section 2. The requirements of a connected smart health framework are presented in Section 3, which is followed by the presentation of a connected smart framework in Section 4. Section 5 then illustrates the collaborative capabilities of the connected smart platform, which is followed by Section 6 discussing the federated data analytics approach . Implementation and validation are presented in Section 7. Finally, Section 8 concludes the article by highlighting ongoing and future work.

## 2 LITERATURE REVIEW

The literature primarily focuses on tackling the convergent research directions addressed in this article independently through related work endeavours. Specifically: *(i) Smart Healthcare Architectures & Platforms*, *(ii) Big data integration and interoperability, (iii) Federated and distributed machine learning in healthcare, and (iv) Data security and privacy in healthcare*. Each of these components is integral to the concept of connected health, which is essential for enhancing patient engagement, facilitating real-time health monitoring, and improving health outcomes [3], [6]. This Section is concluded by a comparative analysis, which identifies the gaps in the literature and appraises the contributions presented in this paper.

### 2.1 Smart Healthcare Architectures & Platforms

Research focusing specifically on smart healthcare platforms and architectures remains limited in literature. Notable contributions in this direction include the introduction of an Internet of Things (IoT)-aware smart architecture designed for the automated surveillance and monitoring of pregnant women [7]. Similarly, in [8] a general approach for consultations has been proposed, aimed at minimizing interactions during pandemic scenarios while reducing resource usage and time consumption, utilizing the Internet of Medical Things and a multi-agent system paradigm.

The study in [9] proposed a cloud-based smart IoT platform designed for personalized healthcare data gathering and monitoring through the integration of various IoT devices and electronic health records. The platform features a user-friendly interface that enables real-time communication between patients and providers, facilitating continuous health tracking and timely interventions. It processes and visualizes data to support informed decision-making and employs security measures by focusing on electronic medical records (EMRs). Furthermore, the system ensures interoperability through a data interoperability module that utilizes platform-neutral data interchange formats.

Moreover, the SM-IoT platform in [10] presents a solution for personalized healthcare monitoring by integrating data from both IoT and non-IoT sources. This integration ensures semantic interoperability through a Data Interoperability module, which employs a semantic annotation model to facilitate the consistent understanding and utilization of medical information. The platform features user-friendly interfaces that addresses usability for both patients and healthcare providers. Additionally, security measures are implemented to protect sensitive information, including contract-based access controls that govern data sharing and the ability for users to define privacy rules. Furthermore, the platform incorporates decision support systems for emergency detection and fosters social networking among stakeholders, enhancing communication and collaboration within the healthcare ecosystem.

In a similar vein, the authors in [11] propose a Patient-Centric Healthcare Framework that enhances interoperability among healthcare systems through a tiered architecture integrating Blockchain, Cloud, and IoT technologies. Interoperability is achieved as Blockchain provides a secure and decentralized data exchange mechanism, ensuring healthcare data is immutable and auditable, while smart contracts automate data sharing and access control. The Cloud facilitates scalable storage and processing power, allowing for the integration of diverse data sources and enabling real-time data access and analysis. IoT collects real-time patient data from various devices, which can be securely transmitted to the Cloud and recorded on the Blockchain. The framework also emphasizes semantic interoperability by standardizing data formats and protocols, enabling different systems to meaningfully interpret shared data.

The authors in [12] develop a blockchain-based framework for electronic health records (EHRs) that addresses interoperability, facilitating data sharing across various healthcare systems. This framework employs a blockchain approach to efficiently and securely gather health data from multiple sources, enabling real-time access and management. It emphasizes interoperability by integrating existing EHR standards, such as HIPAA and HL7, into the blockchain framework, which allows diverse healthcare providers to share and access patient data without reliance on central authorities. Additionally, the framework ensures security through the use of cryptographic techniques, smarts contracts, and decentralized data storage, protecting sensitive health information from unauthorized access and tampering.

While most discussed frameworks/architectures focus on IoT solutions, the approach proposed in this paper prioritizes patient engagement and communication. Furthermore, the collaborative privacy-preserving federated analytics approach discussed in this paper does not only emphasizes on creating an easy-to-use, secure networking platform for all stakeholders in the care continuum, it also preserves the privacy of medical data by employing a virtual data lake, which does not hold or store any datasets of patient data Furthermore, an ontology-driven approach is incorporated to tackle the heterogeneity challenge in big data that enables the synchronization and coordination of chronic care across care teams and locations and provides a big picture. This paves the way for the application of diverse machine learning capabilities for informed data-driven fact-based decision making and actionable intelligence.

## 2.2 Big Data Integration and Interoperability

Data from diverse sources are characterized by multiple types of heterogeneity [13]: (i) syntactic heterogeneity: is a result of differences in representation format of data (models or languages), (ii) schematic or structural heterogeneity: when the structure to store data differ in data sources leading to structural heterogeneity, and (iii) semantic heterogeneity: is caused by different meanings or interpretations of data. Big data challenges in healthcare are intensified due to the dispersion of heterogeneous data among stakeholders. The dispersion of clinical data results in a fragmented view of a patient's record and making treatment delivery less efficient and variable.

The study in [14] addresses both format and semantic interoperability by introducing the MDII-RMHD framework, which enhances interoperability among healthcare devices through the use of data conversion agents that facilitate seamless communication. These agents enable devices to translate data formats, allowing diverse medical devices to communicate. Additionally, the framework integrates cloud-based resources at the network's edge, which reduces reliance on centralized systems and promotes real-time data sharing.

Moreover, the authors in [15] propose a mechanism that integrates both technical and semantic interoperability. Technical interoperability is facilitated through the collection and exchange of data from various IoT medical devices via protocols like Bluetooth Low-Energy (BLE), ensuring compatibility through data cleaning processes. Semantic interoperability is achieved by transforming high-quality data into the HL7 Fast Healthcare Interoperability Resources (FHIR) standard, enabling a shared understanding of the data across different systems. The framework emphasizes the importance of data quality assessment as a prerequisite for interoperability

Data interoperability in the proposed SDHP is achieved by developing an extensible ontology-based approach using OWL2.0 standard that addresses the syntactic, schematic and semantic heterogeneity of big data. This enables: (i)sharing a common understanding of the structure, meaning and context of holistic patient information, among all stakeholders in the care continuum and software agents, (ii) reusing domain knowledge, (iii) making domain assumptions explicit, (iii) separating domain knowledge from the operational knowledge, and (iv) analyzing domain knowledge, as OWL is formally grounded on Description Logic (DL) [16]. Additionally, we adhere to HL7 FHIR[4] standard to ensure compatibility and facilitate seamless data exchange. The details of this data interoperability approach is presented in [[17], and is outside the scope of this article.

## 2.3 Federated and Distributed Machine Learning

The rapid evolution of machine learning has transformed various industries, particularly in the realm of data-driven decision-making. Traditional centralized machine learning approaches, which rely on aggregating large volumes of data in a single location, face significant challenges related to privacy, security, and scalability. In response to these limitations, federated and distributed machine learning have emerged as innovative paradigms that enable decentralized data processing [18], [19].

---

[4] HL7 FHIR release 5: https://www.hl7.org/fhir/

The study in [20] presents a collaborative federated learning framework using clustered federated learning (CFL) for multi-modal COVID-19 diagnosis, leveraging edge computing to enhance data privacy and security. The study in [21] presents OpenFL an open-source framework for federated learning that enables secure decentralized collaboration on machine learning models without sharing sensitive data. Similarly, [22] introduces ADDETECTOR as a privacy-preserving system for early-stage Alzheimer's disease detection that utilizes voice samples collected via IoT devices. It employs federated learning and advanced linguistic and acoustic features to enhance detection accuracy. The system integrates differential privacy mechanisms to protect user information during data transmission.

The proposed framework utilizes a virtual data lake and a central federated learning approach, particularly FedAvg [4], [23] to facilitate efficient model training while maintaining data privacy. This method allows the aggregation of model updates from multiple decentralized sources without transferring the actual data, thereby enhancing security and scalability in health applications.

## 2.4 Data Security and Privacy

The software industry is confronted with the imperative of developing secure healthcare applications that effectively balance usability with robust protection mechanisms [24]. Regulatory compliance, such as GDPR, is critical, especially in contexts where AI is employed.

The study in [25] provides a comprehensive review of the ethical challenges and implications associated with the use of AI in healthcare, specifically within the GDPR framework. It underscores the necessity for transparency, accountability, and the mitigation of bias in AI applications, alongside a thorough consideration of both practical and legal dimensions. The article advocates for stringent data protection laws, the necessity of informed consent, and ethical considerations surrounding the monetization of health data. Ultimately, it posits a rights-based approach aimed at safeguarding individual autonomy while simultaneously enhancing access to quality healthcare services.

Similarly, regulatory compliance frameworks, such as the HIPAA, are explored in study [26]. This research introduces a benchmarking framework that integrates fuzzy logic, Analytic Network Process (FANP), and Technique for Order of Preference by Similarity to Ideal Solution to evaluate usable security within hospital management systems. The study identifies critical security factors, including confidentiality, integrity, and availability, while emphasizing the pressing need to address healthcare data breaches. It advocates for the integration of security measures into the software design process to improve usability and user satisfaction. Notably, while there has been a focus on assessing usability and security through Multi-Criteria Decision Making methods, FANP and TOPSIS in the evaluation of hospital software remains limited.

Additionally, the issue of data decentralization is recognized as a significant security challenge. Although many researchers have explored federated learning, as discussed in Section 2.4, other work stream utilized blockchain for addressing this issue. For example, in [27] a blockchain-based healthcare interoperability framework is proposed, which enhances the secure sharing of patient medical records through the utilization of the Interplanetary File System (IPFS)

In addition to the federated/distributed analytics approach as discussed in Section 2.3, the design and implementation of the privacy-preserving framework and its SDHP proposed in this article have considered the recommendations in [25], [26] in compliance with GDPR. Furthermore, the platform incorporates techniques for: (i) *Multi-Factor Authentication (MFA), (ii) Role-Based Access Control (RBAC), (iii) Input validation, and (iv) Network isolation*, as well as GDPR notice and informed consent.

### 2.5 Comparison

Table 1 appraises the privacy-preserving collaborative SDHP and its collaborative and federated analytical approach presented in this article to prominent research efforts. The comparative study takes into account an elaborate set of evaluation criteria, which include:

- *Requirement Elicitation Approach*: assesses how stakeholder needs are gathered and defined, ensuring the platform meets user requirements and addresses specific healthcare challenges.
- *Collaborative Aspects*: indicates whether collaboration features are supported and its degree of support.
- *Viewpoints Support*: evaluates how well the platform accommodates different perspectives of users, enabling tailored interactions based on individual roles or contexts.
- *Usability*: assesses the ease of use and accessibility of the platform, focusing on user experience and overall functionality.
- *Data Decentralization*: examines whether the platform allows distributed data storage and processing, enhancing control for individual data owners.
- *Data Integration/Interoperability:* refers to the ability of the platform to connect and exchange data between diverse systems, ensuring collective utilization of information.
- *Analytical Capabilities*: evaluates the platform's ability to perform advanced data analysis, including statistical methods and machine learning techniques.
- *Real data support*: assesses the types of data utilized, distinguishing between real-life data from everyday healthcare practices and structured data from randomized controlled trials (RCTs), as well as the diversification of data.

Table 1: Comparison between related work efforts for collaborative healthcare platforms/architectures

| | Requirement Elicitation | Collaborative Aspects | Viewpoints Support | Usability | Analytical Capabilities | Decentralization | Interoperability | Data Security and Privacy | Real Data Support | Extensibility | Standards Support |
|---|---|---|---|---|---|---|---|---|---|---|---|
| [7] | Semi-supported-Uses surveys. | Semi-supported - Real-time monitoring | Supported | Semi-supported -User-friendly mobile app | Semi-Supported- data mining | Not Supported | Semi-Supported- Schematic | Not Supported | Supported—Diversified Real data | Supported. | Not Supported |
| [8] | Not Supported | Not Supported | Not Supported | Semi-supported Focus on user-friendly app. | Not Supported | Not Supported | Semi-Supported - Schematic | Not Supported | Semi Supported - real-life IoT | Semi-supported | Not Supported |
| [9] | Not Supported | Semi-supported - among healthcare providers. | Supported | Semi-supported- GUIs | Not Supported | Not Supported | Supported-Semantic | Semi-supported-encryption algorithms | Semi Supported — real-life IoT | Semi-supported | Not Supported |
| [10] | Not Supported | Semi-supported Patient-Physician | Supported | Semi-supported- User friendly GUIs | Semi-Supported Built-in analytics tools | Supported | Supported-Semantic | Semi-supported-contract-based access protocols | Semi Supported - real-life IoT | Semi-supported | Supported-HL7 |
| [11] | Not Supported | Semi-supported-Group feature collaboration. | Supported | Semi-supported- user experience | Semi-Supported Built-in analytics tools. | Supported-blockchain technology | Supported- Semantic | Semi-supported-blockchain | Not mentioned | Supported. | Supported-HL7 FHIR. |
| [12] | Not Supported. | Not Supported. | Supported | Not Supported. | Not Supported | Supported-blockchain technology | Semi-supported schematic, semantic | Semi-supported-blockchain | Not mentioned | Semi-Supported | Supported-HL7, HIPAA, and DICOM |
| SDHP | Fully supported-Interviews, Continuous Feedback | Fully supported—Group features, real-time collaboration tools (D2D, D2P, P2P) | Supported | Fully Supported- user-friendly GUIs, Cross platform. | Fully supported-Comprehensive Patten-based analytical tools | Supported-Federated learning; decentralized storage; virtual data lake | Fully supported-APIs, Ontology Driven approach | Fully supported-data decentralization, access controls, virtual data lake, GDPR compliance | Fully Supported— Diversified Real data | Supported—Modular design, API-architecture; other domains | Supported-HL7 FHIR. |

- *Extensibility*: refers to the approach ability to be extended and applied on other healthcare domains, and tacking evolving requirements.
- *Data Security/Privacy*: evaluates the measures implemented to protect sensitive health data, addressing privacy concerns and safeguarding against unauthorized access.
- *Standards support*: indicates whether the approach supports healthcare data standards.

Table 1 shows that the privacy-preserving collaborative smart digital framework and its implemented platform presented in this work, as well as its federated analytical approach and technology stack, are significantly more advanced than state-of-the-art efforts that fully meet all the defined comparative criteria.

## 3 REQUIREMENTS OF CLINICIANS AND PATIENTS

To realize the ambitious objectives of the work presented in this article, we have followed a highly iterative/agile system development methodology that started with a structured iterative Requirements Engineering (RE) approach to identify the functional and non-functional requirements of the SDHP and its interacting components.

Table 2 Essential functional and non-functional requirements of clinicians and patients

|  | Clinicians | Patients |
|---|---|---|
| Functional Requirements | **F-Req#1**: network with other clinicians for medical advice and consultation<br>**F-Req#2**: network with other clinicians for sharing their experience.<br>**F-Req#3**: Access to a library of clinical guidelines and recent research to stay updated treatment advances.<br>**F-Req#4**: Coordinate integrated holistic care pathways among all care members, which outlines treatment goals, timelines, and responsibilities for both the physician and the patient.<br>**F-Req#5**: collaborate with their patients in real-time on a regular basis<br>**F-Req#6**: Recommend on treatment outcomes based on patient demographics and medical history.<br>**F-Req#7**: Predict adverse events of a cancer patient receiving/will receive immunotherapy treatment.<br>**F-Req#8**: Recommend a treatment plan for cancer patient with specific inherent characteristics. | **F-Req#9**: connect in a controlled environment with other patients who are receiving/have received immunotherapy to share experience and for emotional support<br>**F-Req#10**: communicate with their doctors in real-time on a regular basis.<br>**F-Req#11**: Understand their treatment options and make a shared informed decision with their physicians.<br>**F-Req#12**: Understand the adverse events of their immunotherapy treatment and make a shared informed decision with her physician.<br>**F-Req#13**: use a mobile application that allows for easy communication with their healthcare team and tracking of their health metrics. |
| NFR | **NF-Req#1**: Ensuring the security and confidentiality and privacy of patient's data<br>**NF-Req#2**: High usability, which includes the ease of use of the SDHP and the user experience. | |

The RE process started with identifying the key stakeholders, which focused on clinicians (from four EU countries that constitute 10 medical institutions) and patients' representatives from the QUALITOP project. RE commenced with extensive requirements elicitation activities through focused and group interviews and collecting active comments and suggestions from the stakeholders. This is followed by defining and modelling the identified requirements using standardized UML models, prioritizing requirements, developing user-interface dialogs to enable the stakeholders to validate the identified requirements and evaluating requirements with users. Table 2 only highlights that core functional and non-functional requirements that the stakeholders have marked as "Essential". The details of this RE process are outside the scope of this article. The next sections discuss how these requirements are realized through a smart digital framework (Section 4) and a supporting smart digital collaborative platform (Sections 5 and 6).

## 4 SMART DIGITAL FRAMEWORK FOR PERSONALIZED PREVENTION AND PATIENT MANAGEMENT

The smart digital framework in this article provides the tools for patient-centered self-management by engaging healthcare professionals and patients in shared informed decision-making while preserving patient privacy, by means of: (i) Smart Digital Health Platform, (ii) virtual datalake, (iii) semantic ontology, and (iv) federated data analytics/machine learning approaches.

Figure 1 presents the privacy-preserving smart digital framework for personalized prevention and patient management. The SDHP, knowledge models and the virtual data lake that can acquire, ingest, and process data from various sources and store the results as a single source of truth of consolidated information in the Data Lake. Knowledge models harmonize different local contextual meta-data embeddings and unify them into a global model. This contextual embedding demonstrates a very strong representative capability to describe medical concepts (and their context), and shows promise as an alternative way to support machine learning applications without the need to disclose original data through federated analytics [17].

Agile software development approach was used to determine the advanced analytical needs of medical institutions in the four EU countries. This process began with a structured Requirements Engineering phase to identify common, recurrent analytical query patterns, which were then incrementally modeled and formalized. Analytical query patterns are expressed against the semantic knowledge models, which are stored in the Data Lake [28], are decomposed and are dispatched to the local databases in parallel. Decomposed analytical queries are executed in the local data sources and the results obtained are synthesized and processed on the SDHP using different pre-built modules to help healthcare executives gain actionable insights.

We use a simple scenario to exemplify the process of federated querying. In this scenario we assume that: *a medical professional in Amsterdam is interested in predicting the adverse events of his female melanoma patient, who is 40-year-old, classified with TNM stage "T3AN2Cm0" and undergoing treatment with "Pembrolizumab: 200 mg" at a frequency of "Q3W"*. Note that this analytical query corresponds to F-Req#7 presented in Section 3.

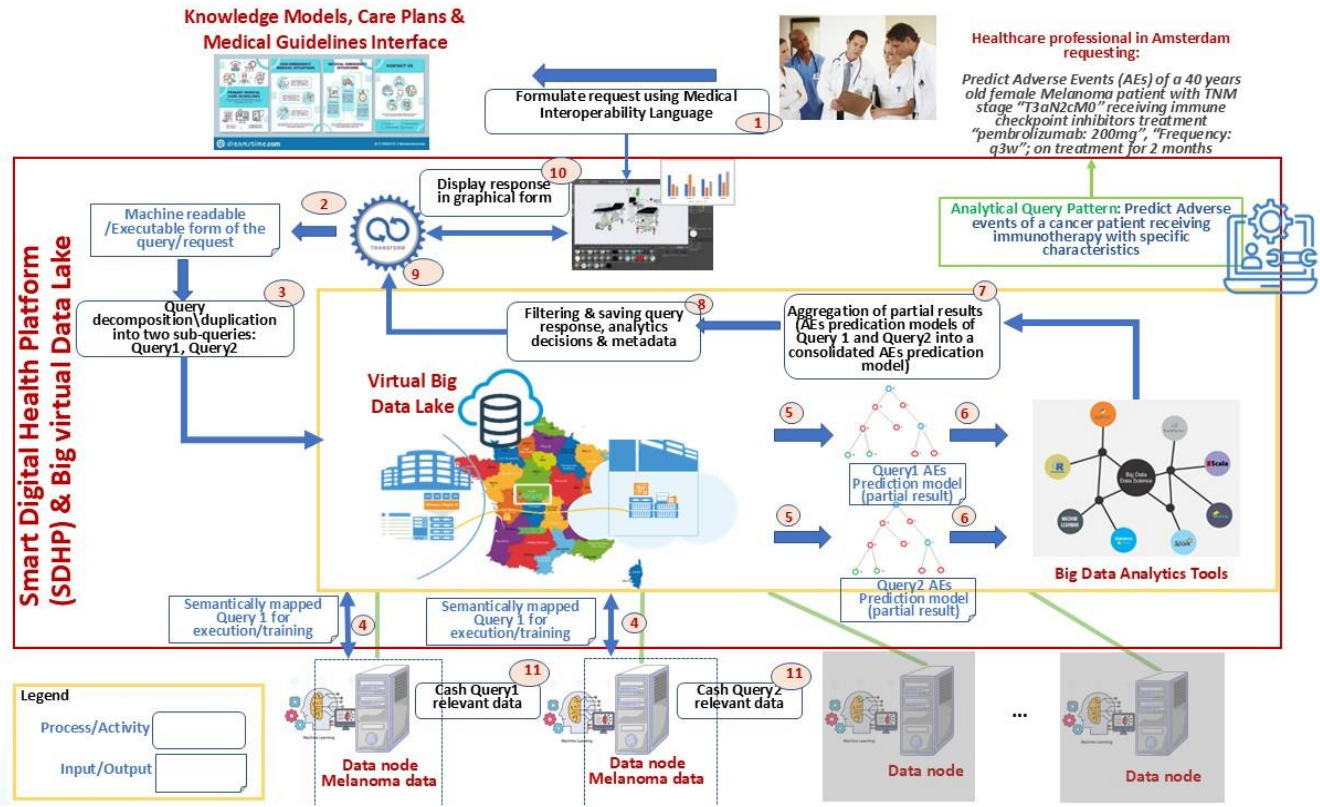

Figure 1 Privacy-preserving smart digital framework for personalized prevention and patient management

In Figure 1 this analytical query request is expressed in a simple natural language – using the proposed user friendly GUIs (realizing NF-Req#2)– expressed against the semantic knowledge models, care plans and guidelines (step (1) in Figure 1), which are all stored in the virtual data lake. This user-friendly high-level request is then transformed into an executable query expressed against the Data Lake by means of a transformation engine (step (2) in Figure 1). The SDHP then analyzes the executable query and generates remote and local access plans (step (3) in Figure 1) that trigger sub-queries against data-node#1 and data-node#2, by assuming that Melanoma data exists in data-node#1 and data-node#2 located in France and Spain, respectively (step (4) in Figure 1). The SDHP consults the data source wrapper in the local and remote data sources and the information stored on a global catalogue in the Data Lake in order to help it process and decompose the federated query into query fragments.

Inside the virtual data lake, each sub-query is translated into the executable language accepted by the respective local data source. A routing component will then forward each sub-query to the respective local data sources for execution, following an edge computing approach. The sub-queries are executed locally, and the query results are returned to the data lake (step (5) in Figure 1), which may need to go through a further analytical step based on the nature of the (analytical) query (step (6) in Figure 1). This is followed by an aggregation component that aggregates the results of the two sub-queries (step (7) in Figure 1). The aggregated analytical result is then filtered and saved in the data lake (step (8) in Figure 1), transformed into a graphical format, e.g., an interactive graph, chart, etc. (step (9) in Figure 1) and displayed on a GUI of the SDHP (step (10) in Figure 1 and realizing NF-Req#2) . Finally, at each local node, the data used for analytics/training may be cashed at the local nodes for model re-training/improvement (step (10) in Figure 1). *The main focus of this article is on: (i) the collaborative smart digital platform and its collaborative capabilities, and (ii) the privacy preserving federated data analytics/machine learning approach. These are discussed in Sections 5 and 6, respectively, which link with the functional and non-functional requirements presented in Section 3.*

## 5  COLLABORATIVE SMART DIGITAL HEALTH PLATFORM

Related to the essential functional and non-functional requirements presented in Section 3, the SDHP realizes all the essential requirements. Timely collaboration and communication capabilities corresponding to functional requirements F-Req#1 to F-Req#5 and F-Req#9 and F-Req#10 are discussed in the following sub-sections by categorizing the communication into: (i) Doctor-to-doctor communication (Section 5.1), (ii) Doctor-to-patient interaction (Section 5.2), and (iii) Patient-to-patient engagement (Section 5.3). NF-Req#1 related to ensuring security and privacy is discussed in Section 2.4. Figure 2 shows a screenshot of the implementation of the SDHP, which is implemented as a secure, private, and user-friendly cross-platform solution that operates seamlessly on both mobile devices (realizing F-Req#13 in Section 3) and as web-based system.

## 5.1 Doctor to Doctor Communication

This feature offers a range of functionalities that facilitate seamless interaction among doctors, including:

- *Q&A Forums*: A dedicated space where doctors can ask questions and share knowledge. Answers can be voted up or down, helping to highlight the best responses and fostering a collaborative environment. This capability realizes F-Req#1.
- *Wiki Pages*: Collaborative documents that allow for the collective creation and editing of medical content. This ensures that information remains current and relevant, benefiting all users. This capability realizes F-Req#1.
- *File Sharing*: A secure method for exchanging important documents, such as research papers, case studies, educational material, and treatment protocols. This enhances collaboration on patient care. This capability realizes F-Req#2 and F-Req#3.
- *Task Management*: Tools that help doctors organize and assign tasks, ensuring accountability and efficient workflow. This capability realizes F-Req#4.
- *Calendar Integration*: A shared calendar that enables doctors to keep track of important dates and deadlines, facilitating better coordination. This capability realizes F-Req#4.
- *Meetings*: Scheduling tools for in-person or virtual meetings to discuss cases or share insights. This capability realizes F-Req#1, F-Req#2 and F-Req#4.
- *Video Conferencing*: A platform for real-time discussions, allowing doctors to consult with one another and enabling integrated care coordination regardless of location. This capability realizes F-Req#1, F-Req#2 and F-Req#4.

## 5.2 Doctor to Patient Interaction

This component provides essential functionalities that enhance communication between doctors and their patients, including:

- *Private Chatting*: A secure messaging feature that allows for direct, confidential communication between doctors and patients. This capability realizes F-Req#5 and F-Req#10.
- *Private Spaces*: Dedicated areas where doctors can share personalized information or resources tailored to individual patients. This capability realizes F-Req#5 and F-Req#10.
- *Appointment Booking*: An easy-to-use system that enables patients to schedule appointments at their convenience, reducing administrative burdens. This capability realizes F-Req#5 and F-Req#10.

## 5.3 Patient to Patient Engagement

This aspect encourages interaction among patients, fostering a supportive community that promotes sharing experience and emotional support. Patients are not allowed to privately chat with other patients:

- *Group Chats*: Spaces where patients can connect, share experiences, and discuss their journeys with others facing similar health challenges. This capability realizes F-Req#10.

- *Virtual Meetings*: Scheduled gatherings that provide patients with opportunities to engage with peers and healthcare professionals in a structured environment. This capability realizes F-Req#9.
- *Support Spaces*: Dedicated forums for emotional and psychological support, helping patients feel less isolated in their experiences. This capability realizes F-Req#9.

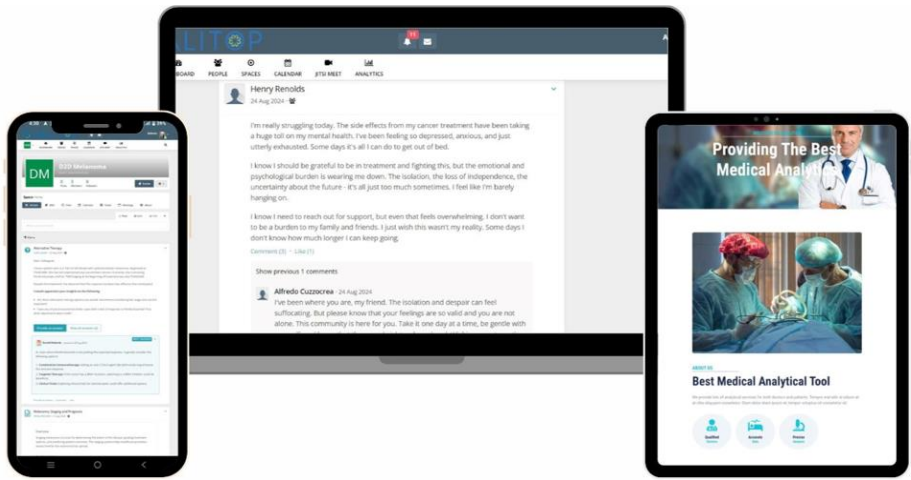

Figure 2 Screenshot of the collaborative smart digital health platform

As a future work direction, Natural Language Processing techniques (NLP) will be employed to ensure that the interactions between patients are only for sharing experiences and for emotional support, and no medical advice are given in such communications.

### 5.4 Privacy and Security

Ensuring the security and privacy of user data is paramount for such a highly-regulated domain, which corresponds to NF-Req#1 in Section 3. To achieve this, we have implemented a comprehensive array of security measures designed to protect sensitive information and facilitate secure interactions among users. Our approach encompasses advanced authentication methods, robust access controls, and proactive threat prevention strategies:

- *Role-Based Access Control (RBAC)*: the SDHP utilizes a robust RBAC system to manage user access. During registration, users select their role (e.g., doctor, patient) and provide their registered email. A confirmation email requests verification of their role, which may involve submitting a doctor's license for physicians or proof of illness for patients. Admins carefully verify these documents to confirm user legitimacy. Once verified, users gain access with permissions that restrict
- *End-to-End Encryption*: the SDHP supports end-to-end encryption for private messages, ensuring that only the sender and recipient can read the messages.

- *Secure Connections*: The platform employs HTTPS to encrypt data in transit, protecting against eavesdropping and man-in-the-middle attacks.
- *Access Controls*: Users can configure privacy settings to enhance their security and manage data visibility.
- *Multi-Factor Authentication (MFA)*: to bolster user account security, the platform incorporates MFA, which adds an extra verification layer beyond the traditional username and password.
- *Time-based One-Time Passwords (TOTP)*: authenticator applications generate time-sensitive codes that users must input with their login credentials.
- *Input Validation*: to defend against potential attacks, the platform implements stringent input validation practices. All user-provided data is thoroughly scrutinized before processing to prevent exploitation through techniques such as SQL injection and cross-site scripting (XSS). This proactive approach ensures that malicious users cannot inject harmful code or exploit vulnerabilities within the platform.

## 6 PRIVACY-PRESERVING FEDERATED DATA ANALYTICS

Federated learning is an innovative decentralized machine learning methodology that facilitates collaborative model training across multiple parties without the necessity of centralizing sensitive data [29]. In this approach, each participant trains a model locally on their own datasets and transmits only model updates to a central server, thereby preserving data privacy and minimizing the risk of breaches. By enabling institutions to collaboratively develop predictive models while retaining data on local systems, federated learning adheres to regulatory frameworks such as GDPR and HIPAA.

Federated learning encompasses various algorithms, each with its strengths and weaknesses. The FedAvg algorithm [30] is frequently deemed appropriate and is attributable to its high degree of simplicity, moderate communication efficiency, and minimal computational requirements. In contrast, the FedProx algorithm [31] demonstrates advantages in effectively managing heterogeneous data distributions. Furthermore, both FedDyn [32] and SCAFFOLD algorithms [33] enhance performance stability across diverse client environments; however, they necessitate greater computational resources. Each algorithm possesses distinct characteristics that can be strategically utilized according to the specific demands of the federated learning application [23]. In the context of the proposed framework, the FedAvg algorithm was selected

### 6.1 The FedAvg Algorithm

The FedAvg algorithm serves as a cornerstone of federated learning by facilitating the aggregation of locally trained models from multiple clients. In this process, each client utilizes its local dataset to train a model and subsequently transmits the updated model parameters to a central server. The server averages these updates to generate a new global model, which is then redis-tributed to the clients for further training. This iterative process continues until the model converges to an optimal solution. The FedAvg algorithm can be mathematically represented as follows:

$$W_T = \frac{1}{N} \sum_{k=0}^{n} W_K^{(T)}$$

where $W_T$ denotes the global model parameters at iteration $T$, N represents the number of clients, and $W_K^{(T)}$ signifies the model parameters from the K[th] client after local training. This formulation illustrates how the global model is updated by averaging the contributions from all participating clients, ensuring a collective improvement while maintaining data privacy.

As discussed in Section 4, recurring analytical patterns were identified. The following are high-priority analytical patterns (analytical patterns are classified into: high, medium and low based on their importance to the medical partners):

- *Model a: Treatment Insights Visualization*: visualizes treatment trends based on patient attributes (e.g., TNM stage, cancer type). The output is a tree structure representing the most likely treatment options for specific patient profiles, providing a quick and interpretable overview of treatment patterns within the data.
- *Model b: Treatment Prediction:* predicts the optimal treatment plan for a patient based on their medical history.
- *Model c: Causation Model for Treatment-Induced Adverse Events (AEs):* which Identifies whether a current adverse event is caused by ongoing treatment.
- *Model d: Predictive Model for Treatment- Induced Adverse Events (AEs):* proactively assess the potential for future AEs associated with a planned treatment.
- *Model e: Adverse Event Type Prediction*: identifies the most likely types of AEs a patient might experience.

In the following, Section 6.2 illustrates the machine learning process at each local node. The "federated learning" process of the machine learning process is then discussed in Section 6.3. This Section is concluded by a performance evaluation of the selected machine learning techniques in Section 6.4

## 6.2 The Machine Learning Process at Each Node

Figure 3 shows the applied machine learning process [34]. This machine learning process was executed at each node in a federated learning setup. The process comprises five sub-processes that are illustrated in the following sub-sections.

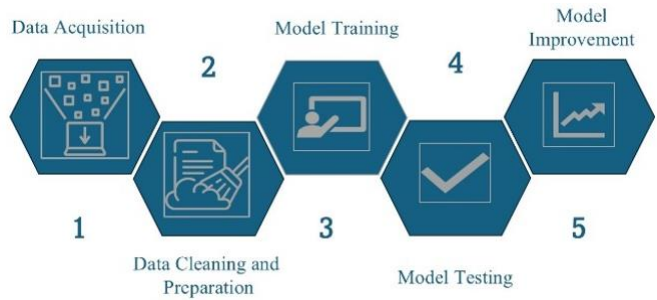

Figure 3 The machine learning process at each node

*6.2.1. Data Acquisition*

In the context of the QUALITOP project, data acquisition process is based on real medical data collected through four Data Transfer Agreements (DTAs)[5] from clinical partners in Portugal, the Netherlands, Spain and France, of patients receiving immunotherapy in real-life. In this cohort study, adult patients are monitored in their real-life, which overcomes the known limitations of Randomized Clinical Trails (RCTs) [35], where patients' profiles and behaviors may be far different from those seen in RCTs. Relevant historical real-world databases and medical administrative registries were also identified. The study included patients recruited specifically for QUALITOP. Patients' clinical health status and QoL are monitored up until 18 months post treatment initiation. Psycho-social wellbeing is also monitored with a tailored QUALITOP questionnaire at baseline and 3, 6, 12 and 18 months post treatment initiation.

*6.2.2. Data Cleaning and Preparation*

This process starts with an "*Inspection and Cleaning*" process, where the datasets are systematically organized into discrete, decentralized nodes. Each node undergoes a comprehensive examination to identify and rectify issues related to missing values, inconsistencies, and duplicate entries. To ensure the integrity of the data, we implemented a series of validation checks, including an assessment of data types for correctness, the execution of summary statistics to detect outliers, and the application of validation rules.

This is followed by "Balancing" process, which evaluates class distribution within each node to identify instances of class imbalance, employing statistical tests such as the Chi-Squared test [36]. Some data sets exhibit significant imbalances, which posed a risk of biased model performance. To mitigate this risk, we employed advanced techniques such as the Synthetic Minority Over-sampling Technique (SMOTE) [37] and Adaptive Synthetic Sampling (ADASYN) [38] to generate synthetic samples for underrepresented classes. This intervention ensured a more equitable distribution of classes within the dataset, thereby enhancing the robustness of subsequent analyses.

---

[5] The QUALITOP cohort study has been registered as an observational study at www.clinicaltrials.gov. Trial registration number NCT05626764

Then "*Transformation*" of data is performed, which revealed that substantial portion of the datasets comprises categorical variables. We utilized one-hot encoding technique [39] to convert these variables into a numerical format conducive to model training. This transformation was essential for integrating categorical data into the selected machine learning algorithms (i.e., Support Vector Machines (SVM), Random Forest, and Decision Trees), thereby facilitating more effective analysis and predictive modeling.

*6.2.3. Semantic Meta Data Mapping*

The process of contextual embedding (metadata) retrieval initiates with the collection of metadata from the distributed data nodes. This metadata encompasses essential information regarding the structure, attributes, and interrelationships present within each data source. Following the aggregation of this metadata, it is mapped to a knowledge base ontology (cf. Section 4), which standardizes the attributes and relationships in accordance with predefined ontology schemas [17]. By mapping the collected metadata to these schemas, we establish a unified representation of the distributed data, thereby facilitating the creation of a virtual view. This virtual view enables users to perceive and interact with the data as if it were a singular, cohesive dataset. Upon receiving a query, the system engages in a two-step mapping process:

1. Mapping the query to original data nodes: the system identifies the original data nodes that contain the actual data pertinent to the query. This step involves determining which segments of the virtual view correspond to specific nodes within the federated system.
2. translating queries to node-specific formats: the query is then translated to conform to the original attribute names and relationships defined within the individual data nodes. This translation ensures that the query accurately retrieves data from each node while adhering to its native schema, structure, and semantics.

*6.2.4. Model Training*

The model training process, as illustrated in Figure 4, outlines the steps followed on each node during the training phase. This process begins with the *Integration of Domain Knowledge*, which is a critical component of our methodology. Collaborating with local medical experts allowed us to enrich each node with clinically relevant insights, thereby enhancing both the model inputs and outputs. By leveraging the expertise of healthcare professionals, we refined our feature selection process, ensuring that the models were not only theoretically sound but also applicable to real-world healthcare scenarios. This integration significantly improved the interpretability of the models, facilitating a better understanding of the clinical implications of the predictions.

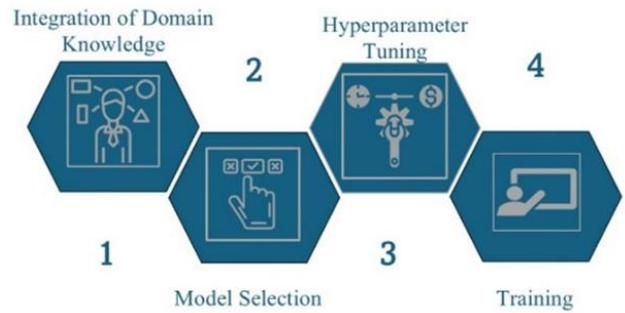

Figure 4 The "Model Training" Process

Following the integration of domain knowledge, the next step is *Model Selection*. This activity focuses on identifying the most appropriate model architecture tailored specifically to the research objectives outlined in this article. The initial exploration encompassed a variety of algorithms, including Support Vector Machines (SVM), Random Forest, and Decision Trees, as shown in Table 3. The selection of SVM was particularly strategic due to its proven efficacy in high-dimensional spaces, which aligns well with the intricate nature of healthcare data. In contrast, Random Forest emerged as a robust choice, demonstrating inherent resilience against overfitting and adeptness at managing mixed data types, thus ensuring comprehensive analysis. Meanwhile, Decision Trees provided a level of interpretability that is crucial in clinical settings, allowing us to trace decision pathways and enhance the transparency of our models. To rigorously determine the optimal model for each specific task, we employed advanced techniques such as cross-validation and grid search [40]. This systematic approach enabled the evaluation of a range of performance metrics, including accuracy, precision, and recall, across various configurations.

The subsequent phase is *Hyperparameter Tuning*, characterized by a combination of randomized search and grid search methodologies aimed at optimizing model performance. This phase involved adjustments to critical parameters, such as the number of trees in the Random Forest, the kernel type in SVM, and the maximum depth of Decision Trees. By systematically evaluating the impact of these adjustments on validation performance, we were able to significantly enhance the predictive capabilities of our models. This iterative refinement not only improved model accuracy but also ensured that the models were robust and adaptable to varying data conditions.

Finally, the *Training phase* involves each node strategically dividing its dataset into training (75-80%) and testing (20-25%) sets to facilitate a thorough assessment of model performance. During this phase, we introduced the prepared data into the models, allowing them to identify and learn from underlying patterns. The models engaged in an iterative process of weight adjustment based on the loss function, which quantified the discrepancy between predicted outcomes and actual results. This continuous feedback loop was crucial for optimizing model performance and ensuring that our predictive models were capable of delivering reliable insights.

*6.2.5. Model Testing*

The initial phase of model testing focused on selecting evaluation metrics that align closely with the specific objectives of the study. We utilized a range of metrics, including accuracy, precision, recall, F1-score, and the area under the receiver operating characteristic curve (AUC-ROC). Given the nature of the data, where class imbalance was a significant concern, we placed particular emphasis on precision and recall. This choice was crucial for ensuring that the models accurately identified relevant clinical outcomes, such as adverse events detection, without generating excessive false positives that could lead to unnecessary interventions. For example, in scenarios where identifying positive cases was critical, high precision ensured that the cases flagged for further investigation were indeed relevant.

Table 3 presents the accuracy of the different algorithms employed during this training phase, with results measured on each local node. These accuracy metrics will be aggregated using the Federated Learning approach (Section 6.3), allowing for a comprehensive evaluation of model performance across all nodes while preserving data privacy.

Table 3 Evaluation of Model Performance at each node

**Accuracy**

| Algo. | Model b | | | Model c | | | Model d | | | Model e | | |
|---|---|---|---|---|---|---|---|---|---|---|---|---|
| | Node #1 | Node #2 | Node #3 | Node #1 | Node #2 | Node #3 | Node #1 | Node #2 | Node #3 | Node #1 | Node #2 | Node #3 |
| RF | 0.601 | 0.65 | 0.612 | 0.99 | 0.995 | 0.995 | 0.69 | 0.73 | 0.717 | 0.72 | 0.73 | 0.746 |
| DT | 0.719 | 0.65 | 0.7 | 0.97 | 0.975 | 0.978 | 0.73 | 0.76 | 0.765 | 0.79 | 0.82 | 0.821 |
| SVM | 0.72 | 0.73 | 0.723 | 0.992 | 0.994 | 0.994 | 0.77 | 0.79 | 0.792 | 0.80 | 0.82 | 0.831 |

*6.2.6. Model Improvement*

This process involved multiple cycles of iterative refinement, where models were adjusted based on performance metrics and user interactions, ensuring continuous enhancement. Feedback from users and stakeholders was actively integrated into the model design, aligning the outcomes with the needs and expectations of both healthcare providers and patients, thereby increasing the models' relevance and effectiveness. Additionally, domain-specific adjustments were made by leveraging the expertise of the teams involved, refining features, parameters, and data inputs to address the unique challenges and requirements of cancer immunotherapy.

**6.3 The Aggregated Federated Learning Model Learning Process**

The Aggregated Federated Learning (FL) model learning process is specifically structured to enable collaborative model training across multiple nodes. This process encompasses several critical steps:

1. *Model Initialization*: as discussed in the previous section the models are initialized and deployed on each local node, allowing for localized training on semantically mapped data.
2. *Local Model Training*: during this phase, nodes update their model parameters based on the outcomes of their local training, which incorporates insights from local medical experts and apply techniques like data balancing and transformation.
3. *Parameter Aggregation*: after local training, the updated parameters from each node are transmitted to the central server. The server aggregates these parameters by averaging them, thus updating the global model. This process is implemented using the Federated Averaging (FedAvg) algorithm, ensuring a balanced contribution from all nodes.
4. *Iterative Training*: The cycle of local model training and parameter aggregation was repeated 200 times until the model converged to a satisfactory level of accuracy.

### 6.4 The Aggregated Federated Learning Models Performance Evaluation

We evaluated the performance of three selected machine learning algorithms—SVM, Random Forest, and Decision Tree—against the selected analytical patterns. Each algorithm was assessed based on its accuracy across the four specific analytical tasks presented in Section 6.1.

Table 4 Evaluation of Federated Learning Model Performance

|  | **Accuracy** | | | |
| --- | --- | --- | --- | --- |
| Algorithm | Model b | Model c | Model d | Model e |
| Random Forest | 0.6207 | 0.9935 | 0.7124 | 0.7320 |
| Decision Tree | 0.6897 | 0.9739 | 0.7516 | 0.8104 |
| SVM | 0.7241 | 0.9935 | 0.7843 | 0.8169 |

As illustrated in Table 4, the SVM model consistently outperformed the other algorithms across all tasks, achieving the highest accuracy rates. These results indicate that the SVM model is the most effective choice for the analytical patterns, particularly in predicting treatment outcomes and adverse events.

### 7 IMPLEMENTATION AND VALIDATION

The implementation[6] of the privacy-preserving smart digital health framework presented in this article (cf. Figure 1) is a demanding but necessary endeavor to verify the validity and reliability of the proposed approach. An agile/iterative system development methodology is adopted, which mandates the heavy involvement of system's stakeholders (in the QUALITOP project) in the requirements engineering, design of systems components, implementation, testing and evaluation processes. These involves the continuous validation of the activities of each process that triggered various cycles of refinement and improvement.

---

[6] Video demonstrations of the platform and its analytical module are available at:
https://drive.google.com/drive/folders/1BmFfh2C1yEB5PJeCX0sdgxDdpydsEgKg?usp=sharing

Figure 5 presents a standard UML component diagram showing the overall system architecture and its logical components. The identified and implemented components interact with each other via Application Programming Interfaces (APIs). The implementation comprises of four main components: (i) the SDHP, (ii) Data Homogenization component, (iii) Federated Query Processing component, and (iv) Distributed Hospitals Database System components. In the following discussion (Sections 7.1- 7.4), each component is discussed by showing its interactions with other components. This Section is concluded by the iterative validation and evaluation efforts in Section 7.5.

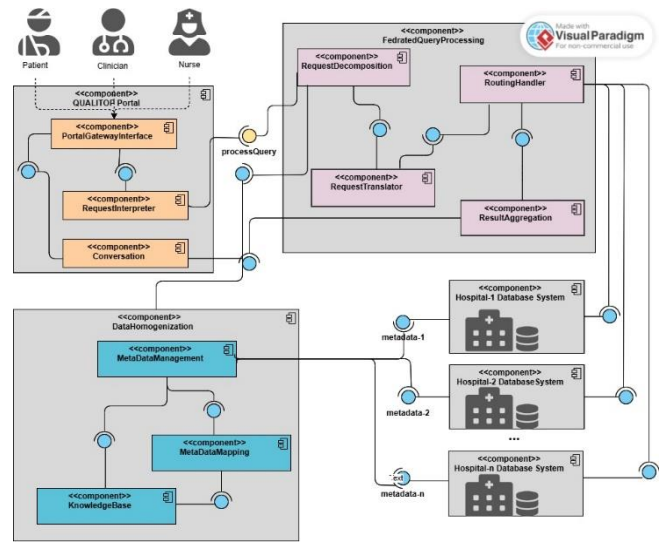

Figure 5 Component diagram of the implementation of the smart digital framework

### 7.1 Data Homogenization Component

This component represents the backbone of the whole system. It consists of:

*Knowledge Base:* implementing the knowledge models discussed in Section 4, as an ontology in the Ontology Web Language[7] (OWL) standard using protégé[8] framework. The details of the OWL implementation is reported in [17].

*MetaDataManagement* that requests and imports meta-data from different involved Hospital DataBase System(s). Such technical/structural meta-data includes information about tables, columns, indexes, constraints, and any other relevant details in the disparate databases and is asso-ciated with the relevant ontology models so that they can be discovered in their respective databases. It also requests the "MetaDataMapping" component to carry out the meta-data mapping of the local databases against the ontology models schema, which acts as a global view over the local database systems. This leads to simple query processing strategies. The cross-schema meta-data mapping is then stored and maintained by the "MetaDataManagement" component.

---
[7] OWL: www.w3.org/OWL/
[8] https://protege.stanford.edu/

### 7.2 Smart Digital Health Platform

The SDHP is designed and implemented as a cross, private collaborative platform portal using Humhub[9], a free and open-source social networking software built on the Yii PHP framework and PHP for back-end implementation (shown in Figure 5 as "QUALITOP Portal"). This component comprises of three components:

*SDHP Gateway Interface*: this component encompasses all graphical user interfaces (GUIs) that enable seamless interaction between users and the system. It allows users to submit analytical queries and manipulate the results effectively. The design of the *gateway* interface prioritizes user experience, ensuring that users can navigate the system intuitively while accessing necessary analytical tools. This interface is implemented using HTML, CSS, and JavaScript, leveraging the Flask framework for integration with the backend.

*Request Interpreter*: this component is responsible for managing communication with the Federated Query Processing component. The Request Interpreter receives user-submitted queries and translates them into SPARQ[10] (SPARQL Protocol and RDF Query Language), which serves as the standard query language and protocol for querying data stored in OWL (Web Ontology Language) ontology models. The implementation utilizes Python for developing the translation logic, ensuring efficient processing of user queries.

*Conversation Component*: following the processing of queries, the results are visualized and presented on integrated dashboards. This component employs various graphical representations, including graphs, charts, and tables, to facilitate the interpretation of data. This component utilizes Python libraries such as Matplotlib, Seaborn, and Plotly for data visualization, ensuring dynamic and interactive presentations of the results. These visualizations not only enhance user comprehension but also support data-driven decision-making processes. Figure 6 illustrates the visualized output for the running scenario described in Section 4. Additionally, the analytical tool embedded within the SDHP, is implemented using the Flask[11] framework for frontend development, while Python is utilized for backend operations. This combination of technologies provides a flexible and scalable environment for data analysis and user interaction. The SDHP is hosted on PythonAnywhere[12].

---

[9] HumHub: https://www.humhub.com/en
[10] SPARQL 1.1 Query Language: https://www.w3.org/TR/sparql11-query/
[11] Flask framework: https://flask.palletsprojects.com/en/3.0.x/
[12] PythonAnywhere: https://www.pythonanywhere.com/

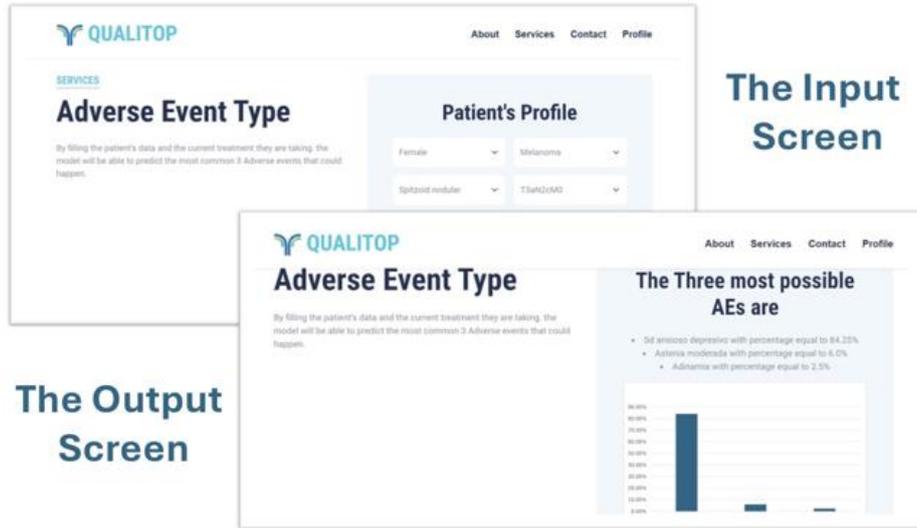

Figure 6 Input and Output of "Adverse Event Type Prediction" analytical pattern

### 7.3 Federated Query Processing Component

The Federated Query Processing Component is a critical element of the Smart Digital Health Platform (SDHP), designed to efficiently manage and execute analytical queries across multiple data sources. This component is structured into several sub-components, each fulfilling specific roles in the query processing workflow.

The *"Request Decomposition"* sub-component receives the SPARQL representation of the analytical pattern from the "Request Translator" sub-component of the SDHP. Implemented in Python, this sub-component utilizes the Flask framework for web integration. Its primary function is to decompose the incoming SPARQL query into multiple sub-queries that correspond to the physical locations of the queried data, such as Hospital-1, Hospital-2, ..., Hospital-n. To achieve this, the "Request Decomposition" sub-component consults the "Data Homogenization" component for metadata mapping, ensuring that the decomposition process aligns with the structure and semantics of the underlying data sources.

The *"Request Interpreter"* sub-component also implemented in Python and utilizing Flask, is responsible for translating each sub-query into the query language accepted by the local database system, e.g., SQL.

The *"Routing Handler"* is then responsible for the execution of the sub-queries by requesting the local database systems of the hospitals to process each respective SQL sub-query locally. This component is implemented in Python and employs (APIs) to facilitate communication with the data nodes. The use of APIs ensures effective data retrieval and enhances the interoperability of the system across different hospital databases.

The *"ResultAggregation"* sub-component aggregating and performing any further processing on the sub-queries results, and then forwarding the aggregated results to the "Conversation" sub-component of the "Portal Gateway" component. The "ResultAggregation" also sends the aggregated query result to the "DataHomogenization" component to be saved and maintained. Python is used for the implementation of this component.

### 7.4 Distributed Hospitals Database System Components

The implementation architecture in Figure 5 is designed as a federation of data products. These data products expose standardized APIs that enable other data sources to access and utilize their data efficiently. This architecture ensures the autonomy of participating hospitals while preserving the privacy of medical data. APIs play a crucial role in various functionalities of the distributed hospital database system, which includes:

*Metadata Requests*: APIs are utilized to request metadata from the participating hospitals. This metadata includes essential information about the data structures, types, and relationships within each hospital's database. By standardizing metadata access, the APIs facilitate seamless integration and interoperability among different data sources.

*Mapping Requests*: APIs are also employed to request data mapping between different databases. This mapping process is vital for aligning data formats and meanings that ensures the consistency across the federated system. The APIs help automate the mapping process, allowing for efficient communication between hospitals and reducing the potential for errors.

*Accessing Mapped Data in Virtual Views*: Once the mapping is established, APIs are used to retrieve mapped data presented in virtual views. These virtual views allow users to query and interact with data from multiple hospitals as if it were a single cohesive dataset. This capability enhances data accessibility and usability, enabling more comprehensive analyses.

*Storing and Updating Model Logs*: APIs facilitate the storage and updating of model logs, which track the interactions and changes made within the database system. This logging is critical for maintaining data integrity, auditing purposes, and ensuring compliance with regulatory requirements. The APIs provide endpoints for logging actions, making it easier to manage and monitor data usage across the system.

By leveraging APIs, the distributed hospitals database system ensures robust, secure, and efficient communication among its components. This architectural choice not only enhances data sharing capabilities but also reinforces the system's commitment to maintaining the privacy and autonomy of each participating hospital.

### 7.5 Validation

The Smart Digital Health Platform (SDHP) is fundamental to the connected smart health framework discussed in this article. It is constructed on a robust framework of sophisticated data analytics, artificial intelligence, and federated learning, having undergone extensive evaluation and refinement during its development phase. This iterative process has involved significant collaboration with medical professionals and representatives from patient advocacy organizations, utilizing the platform in practical applications and engaging in discussions within the context of the QUALITOP project. These collaborative efforts exemplify an agile and iterative approach to system development, facilitating a continuous feedback mechanism that ensures the platform aligns with the needs and expectations of its users.

Through regular online and in-person meetings throughout the duration of the QUALITOP project, medical partners and patient representatives have:

- Confirmed the conceptual and technical applicability and effectiveness of the collaborative and analytical capabilities of the SDHP, which have undergone ongoing cycles of refinement and enhancement.
- Continuously identified and validated the descriptive and analytical query capabilities outlined in this article.
- Tested and validated the SDHP as a user-friendly collaborative healthcare system. Usability, a critical aspect of contemporary systems development, assesses how easily users can achieve their objectives when utilizing a service. This has been evaluated through customer satisfaction research methodologies, as affirmed by the medical partners.
- Evaluated the functional capabilities of the SDHP, which provides essential insights regarding the practical applicability of the platform's features and functionalities. This iterative feedback cycle has led to ongoing improvements, ensuring that the SDHP effectively addresses the complexities associated with personalized care management of cancer patients receiving immunotherapy.
- Verified the privacy-preserving features that are integral to the SDHP's design. Compliance with regulatory standards, particularly the General Data Protection Regulation (GDPR), has been a primary focus throughout the development process. Consultations with Data Protection Officers from participating institutions are ongoing to validate that the privacy measures implemented within the SDHP comply with legal requirements, thereby fostering user trust. A patient representative highlighted the importance of ethical oversight in this context.
- Engaged in model refinement, with regular cycles of training and validation for the models, incorporating feedback from healthcare professionals involved in the QUALITOP project. This collaborative process has ensured that the machine learning models are aligned with real-world clinical requirements and can adapt to changing patient care scenarios.

Nevertheless, it has been raised that the platform should be piloted with actual patients in a controlled setting to collect feedback and make necessary adjustments. Discussions have commenced with participating hospitals across Europe to initiate this experimentation.

## 8 CONCLUSIONS AND FUTURE WORK

This article proposes a connected smart healthcare framework for personalized prevention and patient management of cancer immunotherapy patients. The framework is patient-centered by providing the required tools and services, ensuring that the patient receives care in the most proactive and effective manner. The framework is realized by a smart digital platform supporting comprehensive collaborative features to enable all stakeholders in the care continuum to be connected by means of timely exchange and presentation of accurate and useful patient information. The framework utilizes federated big data analytics and artificial intelligence, enabling better insights for informed and shared decision making, while ensuring privacy and security. Identified analytical query patterns are implemented on patients in real-life in four EU countries (France, The Netherlands, Spain and Portugal) and showed an approximate accuracy of 70-90%. This overcomes the known limitations of Randomized Clinical Trails, where patients' profiles and behaviors may be far different from those seen in RCTs.

The validation and evaluation of the proposed approach is ascertained by its full implementation, the usage of real-life data for the implementation of the federated data analytics approach, and the continuous closed feedback-loop evaluation with medical partners and patients' representatives from the QUALITOP project.

Future work is focusing on: (i) the utilization of blockchain technology to help health information exchanges (HIEs) by relieving security concerns, (ii) the incorporation of virtual and augmented reality technologies, and (iii) experimenting the proposed approach with colorectal cancer screening as part of the ONCOSCREEN[13] project.

**AUTHORS' CONTRIBUTION**

Michael P. Papazoglou was instrumental in the conceptualizing Collaborative Smart framework and the interoperability approach. Bernd J. Krämer and Amal Elgammal worked alongside Papazoglou in these conceptualization efforts. Neamat EL-Tazi advised on the presented ideas. Mira Raheem concentrated on the conceptualization of the analytical query patterns, and data engineering, and was responsible for implementing all components of the digital platform, including its collaborative and federated analytics capabilities. Throughout the project's duration, Amal and Mira collaborated closely with medical partners in the iterative requirement engineering process, continuously validating the concepts and developed solutions and wrote this article.

**CONFLICT OF INTEREST**

The authors acknowledge that there is competing financial or non-financial interests in relationship to the work and results presented in this article.

**FUNDING**

This research is partially funded by the EC Horizon 2020 project QUALITOP, under contract number H2020 - SC1-DTH-01-2019 – 875171.

**ACKNOWLEDGMENTS**

We wish to thank Hospital Clinic Barcelona (IDIBAPS), Hospices Civils de Lyon (HCL), University Medical Center Groningen (UMCG), and Instituto Português de Oncologia, Lisboa

---

[13] ONCOSCREEN project: https://cordis.europa.eu/project/id/101097036

(IPOL) for providing the essentials of the pilot study in this paper, and for the continuous evaluation and validation of the reported results. Special thanks goes to Dr Roxana Albu, Chief Scientific Officer, Association of European Cancer Leagues (ECL), and Dr Menia Koukougianni, Fellow of the European Patients' Academy, for their insightful feedback in evaluating and validating the work presented in this paper.# REFERENCES

[1] "Caulfield, B.M. and S.C. Donnelly, What is Connected Health and why will it change your practice? QJM: An International Journal of Medicine, 2013. 106(8): p. 703-707."

[2] C. S. Pattichis and A. S. Panayides, "Connected Health," *Front Digit Health*, vol. 1, 2019, doi: 10.3389/FDGTH.2019.00001/XML/NLM.

[3] K. Colorafi, "Connected health: a review of the literature," *Mhealth*, vol. 2, pp. 13–13, Apr. 2016, doi: 10.21037/MHEALTH.2016.03.09.

[4] A. Elgammal and B. J. Krämer, "A Reference Architecture for Smart Digital Platform for Personalized Prevention and Patient Management," vol. 12521 LNCS, pp. 88–99, 2021, doi: 10.1007/978-3-030-73203-5_7/FIGURES/1.

[5] P. C. Vinke *et al.*, "Monitoring multidimensional aspects of quality of life after cancer immunotherapy: protocol for the international multicentre, observational QUALITOP cohort study," *BMJ Open*, vol. 13, no. 4, Apr. 2023, doi: 10.1136/BMJOPEN-2022-069090.

[6] N. Carroll, M. Travers, and I. Richardson, "Evaluating multiple perspectives of a connected health ecosystem," *HEALTHINF 2016 - 9th International Conference on Health Informatics,* pp. 17–27, 2016, doi: 10.5220/0005623300170027.

[7] X. Li, Y. Lu, X. Fu, and Y. Qi, "Building the Internet of Things platform for smart maternal healthcare services with wearable devices and cloud computing," *Future Generation Computer Systems*, vol. 118, pp. 282–296, May 2021, doi: 10.1016/j.future.2021.01.016.

[8] S. Kouah, A. Ababsa, and I. Kitouni, "Internet of Things Based Smart Healthcare System," in *Lecture Notes in Networks and Systems*, 2023, pp. 267–281. doi: 10.1007/978-3-031-44097-7_29.

[9] S. S. Vellela, V. L. Reddy, D. Roja, G. R. Rao, S. K. Khader Basha, and K. K. Kumar, "A Cloud-Based Smart IoT Platform for Personalized Healthcare Data Gathering and Monitoring System," in *2023 3rd Asian Conference on Innovation in Technology,* 2023. doi: 10.1109/ASIANCON58793.2023.10270407.

[10] A. Dridi, S. Sassi, and S. Faiz, "A smart iot platform for personalized healthcare monitoring using semantic technologies," in *Proceedings - International Conference on Tools with Artificial Intelligence*, Jul. 2017, pp. 1198–1203. doi: 10.1109/ICTAI.2017.00182.

[11] A. N. Gohar, S. A. Abdelmawgoud, and M. S. Farhan, "A Patient-Centric Healthcare Framework Reference Architecture for Better Semantic Interoperability Based on Blockchain, Cloud, and IoT," *IEEE Access*, vol. 10, pp. 92137–92157, 2022, doi: 10.1109/ACCESS.2022.3202902.

[12] F. A. Reegu *et al.*, "Blockchain-Based Framework for Interoperable Electronic Health Records for an Improved Healthcare System," *Sustainability (Switzerland)*, vol. 15, no. 8, Apr. 2023, doi: 10.3390/su15086337.

[13] R. Pastorino *et al.*, "Benefits and challenges of Big Data in healthcare: An overview of the European initiatives," *Eur J Public Health*, vol. 29, pp. 23–27, Oct. 2019, doi: 10.1093/eurpub/ckz168.

[14] Dr. K. Malathi, S. S. Nair, N. Madhumitha, S. Sreelakshmi, U. Sathya, and M. S. Priya, "Medical Data Integration and Interoperability through Remote Monitoring of Healthcare Devices," *J Wirel Mob Netw Ubiquitous Comput Dependable Appl*, vol. 15, no. 2, pp. 60–72, Jun. 2022, doi: 10.58346/jowua.2024.i2.005.

[15] A. Mavrogiorgou, A. Kiourtis, K. Perakis, S. Pitsios, and D. Kyriazis, "IoT in Healthcare: Achieving Interoperability of High-Quality Data Acquired by IoT Medical Devices," *Sensors (Basel)*, vol. 19, no. 9, Apr. 2019, doi: 10.3390/s19091978.

[16] M. Krötzsch, F. Simančík, and I. Horrocks, "A Description Logic Primer," 2013.

[17] Michael P. Papazoglou, Bernd J. Krämer, Mira Raheem, and Amal ElGammal, "Medical Digital Twins for Personalized Chronic Care," *pre-print, Research Square, Under consideration at Journal of Healthcare Informatics Research*.

[18] M. Joshi, A. Pal, and M. Sankarasubbu, "40 Federated Learning for Healthcare Domain-Pipeline, Applications and Challenges," 2022, doi: 10.1145/3533708.29

[19]     D. C. Nguyen *et al.*, "Federated Learning for Smart Healthcare: A Survey," *ACM Computing Surveys (CSUR)*, vol. 55, no. 3, Feb. 2022, doi: 10.1145/3501296.

[20]     A. Qayyum, K. Ahmad, M. A. Ahsan, A. Al-Fuqaha, and J. Qadir, "Collaborative Federated Learning for Healthcare: Multi-Modal COVID-19 Diagnosis at the Edge", doi: 10.1109/OJCS.2022.3206407.

[21]     P. Foley *et al.*, "OpenFL: the open federated learning library," *Phys Med Biol*, vol. 67, no. 21, Nov. 2022, doi: 10.1088/1361-6560/ac97d9.

[22]     J. Li *et al.*, "A Federated Learning Based Privacy-Preserving Smart Healthcare System," *IEEE Trans Industr Inform*, vol. 18, no. 3, pp. 2021–2031, Mar. 2022, doi: 10.1109/TII.2021.3098010.

[23]     A. Nilsson, S. Smith, G. Ulm, E. Gustavsson, and M. Jirstrand, "A performance evaluation of federated learning algorithms," *DIDL 2018 - Proceedings of the 2nd Workshop on Distributed Infrastructures for Deep Learning*, pp. 1–8, Dec. 2018, doi: 10.1145/3286490.3286559.

[24]     A. I. Newaz, A. Kumar Sikder, M. A. Rahman, and A. Selcuk Uluagac, "A survey on security and privacy issues in modern healthcare systems: Attacks and defenses," *dl.acm.org*, vol. 2, no. 3, Jul. 2021, doi: 10.1145/3453176.

[25]     M. Amini *et al.*, "Artificial intelligence ethics and challenges in healthcare applications: a comprehensive review in the context of the European GDPR mandate," *mdpi.com*, vol. 5, pp. 1023–1035, 2023, doi: 10.3390/make5030053.

[26]     R. Kumar, M. T. Jamal Ansari, A. Baz, H. Alhakami, A. Agrawal, and R. A. Khan, "A multi-perspective benchmarking framework for estimating usable-security of hospital management system software based on fuzzy logic, ANP and TOPSIS methods," *KSII Transactions on Internet and Information Systems*, vol. 15, no. 1, pp. 240–263, Jan. 2021, doi: 10.3837/TIIS.2021.01.014.

[27]     R. Kumar, "Scalable Inter-operable and Secure Healthcare Framework For Sharing Patient Medical Report using Blockchain and IPFS Technology," Oct. 2022, doi: 10.21203/RS.3.RS-2115239/V1.

[28]     C. Giebler, C. Gröger, E. Hoos, and H. Schwarz, "Leveraging the Data Lake Current State and Challenges Institute for Parallel and Distributed Systems / AS," 2019, doi: 10.1007/978-3-030-27520-4_13.

[29]     A. Rahman *et al.*, "Federated learning-based AI approaches in smart healthcare: concepts, taxonomies, challenges and open issues," *Cluster Comput*, vol. 26, no. 4, pp. 2271–2311, Aug. 2023, doi: 10.1007/s10586-022-03658-4.

[30]     H. Brendan McMahan, E. Moore, D. Ramage, S. Hampson, and B. Agüera y Arcas, "Communication-Efficient Learning of Deep Networks from Decentralized Data," *Proceedings of the 20th International Conference on Artificial Intelligence and Statistics*, Feb. 2016, [Online]. Available: https://arxiv.org/abs/1602.05629v4

[31]     T. Li, A. K. Sahu, M. Zaheer, M. Sanjabi, A. Talwalkar, and V. Smith, "Federated Optimization in Heterogeneous Networks," *Proceedings of Machine Learning and Systems*, vol. 2, pp. 429–450, Mar. 2020.

[32]     C. Jin, X. Chen, Y. Gu, and Q. Li, "FedDyn: A dynamic and efficient federated distillation approach on Recommender System," *Proceedings of the International Conference on Parallel and Distributed Systems*, vol. 2023-January, pp. 786–793, 2023, doi: 10.1109/ICPADS56603.2022.00107.

[33]     I. Mandric, J. Lindsay, I. I. Măndoiu, and A. Zelikovsky, "Scaffolding Algorithms," *Computational Methods for Next Generation Sequencing Data Analysis*, pp. 105–131, Sep. 2016, doi: 10.1002/9781119272182.CH5.

[34]     V. Gupta, V. K. Mishra, P. Singhal, and A. Kumar, "An Overview of Supervised Machine Learning Algorithm," *Proceedings of the 2022 11th International Conference on System Modeling and Advancement in Research Trends*, pp. 87–92, 2022, doi: 10.1109/SMART55829.2022.10047618.

[35]     P. C. Vinke *et al.*, "Monitoring multidimensional aspects of quality of life after cancer immunotherapy: protocol for the international multicentre, observational QUALITOP cohort study," *BMJ Open*, vol. 13, no. 4, Apr. 2023, doi: 10.1136/BMJOPEN-2022-069090.

[36]     M. L. McHugh, "The Chi-square test of independence," *Biochem Med (Zagreb)*, vol. 23, no. 2, pp. 143–149, Jun. 2013, doi: 10.11613/BM.2013.018.

[37]     N. V. Chawla, K. W. Bowyer, L. O. Hall, and W. P. Kegelmeyer, "SMOTE: Synthetic Minority Over-sampling Technique," *Journal of Artificial Intelligence Research*, vol. 16, pp. 321–357, Jun. 2002, doi: 10.1613/JAIR.953.

[38]     H. He, Y. Bai, E. A. Garcia, and S. Li, "ADASYN: Adaptive synthetic sampling approach for imbalanced learning," *Proceedings of the International Joint Conference on Neural Networks*, pp. 1322–1328, 2008, doi: 10.1109/IJCNN.2008.4633969.

[39]     C. Seger, "An investigation of categorical variable encoding techniques in machine learning: binary versus one-hot and feature hashing," 2018.

[40]     D. Berrar, "Cross-validation," *Encyclopedia of Bioinformatics and Computational Biology: ABC of Bioinformatics*, vol. 1–3, pp. 542–545, Jan. 2018, doi: 10.1016/B978-0-12-809633-8.20349-X.